\documentclass{iopart}
\usepackage{color}
\usepackage{setstack}
\usepackage{iopams}
\usepackage{epsf}
\usepackage{epsfig}
\usepackage{amssymb}
\usepackage{ulem}

\begin{document}

\title[Masses and Majorana fermions in graphene]
{Masses and Majorana fermions in graphene}
\vskip 10pt
\author{
Claudio Chamon$^{1}$, 
Chang-Yu Hou$^{2}$, 
Christopher Mudry$^{3}$, 
Shinsei Ryu$^{4}$,
Luiz Santos$^{5}$
}
\address{
$^1$ Physics Department, Boston University, 
590 Commonwealth Avenue, Boston, MA 02215, USA
\\
$^2$ Instituut-Lorentz, Universiteit Leiden,
P.O. Box 9506, 2300 RA Leiden, The Netherlands
\\
$^3$ Condensed Matter Theory Group, Paul Scherrer Institute, 
CH-5232 Villigen PSI, Switzerland
\\
$^4$ Department of Physics, University of California, Berkeley, 
California 94720, USA
\\
$^5$ Department of Physics, Harvard University, 
17 Oxford Street, Cambridge, Massachusetts 02138, USA
\\
{\tt chamon@bu.edu, 
hou@lorentz.leidenuniv.nl,
christopher.mudry@psi.ch,
sryu@berkeley.edu,
santos@physics.harvard.edu}
}
\date\today

\begin{abstract}


  We review the classification of all the 36 possible gap-opening
  instabilities in graphene, i.e., the 36 relativistic
  masses of the two-dimensional
  Dirac Hamiltonian when the spin, valley, and superconducting
  channels are included. We then show that in graphene it is possible
  to realize an odd number of Majorana fermions attached to vortices
  in superconducting order parameters if a proper hierarchy of mass
  scales is in place.

\end{abstract}

\newpage

\section{Introduction}
\label{sec:intro}

The discovery~\cite{Novoselov04} that it is possible to peel and isolate
individual atomic layers of graphite, i.e., graphene, has led to
an explosion of experimental and theoretical works and ideas from
exotic physics to real material applications~\cite{graphene-review}. 
Graphene is a material with remarkable physical properties, many of
which are consequence of its band structure
at charge neutrality, which is characterized by two Fermi
points in the Brillouin zone. 
An excitation around those Fermi points, 
being linearly proportional to its momentum,
resembles the
dispersion relation of massless relativistic particles.
The low energy theory is then well described by a 
relativistic Dirac Hamiltonian in two dimensions%
~\cite{Wallace47}. 
Massless Dirac
fermions describe graphene, but one may wonder what kinds of mass gaps
could be induced. This question is rather important for device
applications, as the presence of a gap would make graphene behave as a
semiconductor, just like silicon. But the question is also interesting
for fundamental physics reasons, as we discuss below.

A mass in the Dirac equation can be viewed as an order parameter, a
bosonic field. If the mass is generated by spontaneous breaking of a
symmetry -- the Higgs mechanism in graphene -- there can be
spatio-temporal fluctuations in the mass order parameter. In
particular, there could be topological defects in the order parameter:
domain walls, vortices, or hedgehogs, for instance. Alternatively, the
masses could be induced externally, for example if the gaps are
attained by placing graphene on a certain substrate, such as one
where
there is a difference in the potential seen by the two atoms in
the unit cell of graphene, 
or if the substrate is a superconductor. Topological
defects could also be present in this case where the gaps are
externally induced.

Topological defects in an
order parameter can lead to zero modes in
the fermionic spectrum, i.e., zero energy solutions lying in the
middle of the gap~\cite{Jackiw81}. The physical consequences of these
zero modes are rather remarkable. When charge is
conserved, the zero modes bind a fraction of an electron
charge~\cite{Hou07}, while in superconducting systems, they bind
charge neutral Majorana fermions~\cite{Ghaemi-Wilczek}.

In this paper, we classify all possible 36 competing orders of a
Dirac Hamiltonian represented by 16-dimensional Dirac matrices that
encode the spin, valley, and superconducting channels. We also discuss
the simpler cases where only spin and valley degrees of freedom (no
superconductivity), or valley alone (spinless electrons) are
considered. These simpler cases serve as a warm up exercise in gaining
familiarity with the classification construction as the build up of
the increasingly larger representations is carried out.

We then show how this classification can be applied to the problem of
selecting an odd number of Majorana fermions to bind to superconducting
vortices. That the number of zero modes must be odd is important for
applications to topological quantum computing, that we shall discuss
in more detail below. It is possible to achieve an odd number of
Majorana fermions using surface states of topological insulators
~\cite{Fu-Kane}.
But,
naively, Majorana zero modes bound to superconducting vortices in
graphene, first found by Ghaemi and Wilczek~\cite{Ghaemi-Wilczek} in
this context, would come in quadruplets because of the valley and spin
degrees of freedom. We show otherwise. 
If there is a proper hierarchy of mass scales, 
one can tune selectively the number of
Majorana zero modes in graphene from $4\to 3\to 2\to 1\to 0$.

We aim in this paper at a pedagogical description of the mass
classification scheme, and a brief discussion of how the number of
zero modes can be tuned by changing the relative strength of
masses. The original classification of masses in graphene was
presented in ~\cite{Ryu09}, while the tuning of the number of Majorana
modes was carried out in ~\cite{Santos10}. We point the reader to
those papers for details beyond this review.

\section{Classification of masses in graphene}
\label{sec:masses}

Let us first of all define what we mean by a mass in graphene. Say we
take generically a two-dimensional Dirac Hamiltonian
\begin{equation}
  \label{eq:Dirac-1}
  H=p^{\ }_x\;\alpha^{\ }_1+p^{\ }_y\;\alpha^{\ }_2+m\;M
\end{equation}
where $p^{\ }_{x,y}$ stand for momentum operators in the $x,y$-directions,
and $\alpha^{\ }_{1,2}$ are some (generically) $D$-dimensional matrices
satisfying $\{\alpha^{\ }_i,\alpha^{\ }_j\}=2\,\delta^{\ }_{ij}$. We would like to
say that $m$ is a mass scale, and that $M$ is a mass matrix. For
that to be the case, the matrix $M$ has to satisfy certain commutation
relations with the matrices $\alpha^{\ }_{1,2}$. The relations are seen if
we square the Dirac-type Hamiltonian:
\begin{eqnarray}
  \label{eq:Dirac-2}
  H^2
&=&p^2_x\;+p^2_y\;+m^2\;M^2
+
m\,p^{\ }_x\, \{\alpha^{\ }_1,M\}
+
m\,p^{\ }_y\, \{\alpha^{\ }_2,M\},
\end{eqnarray}
which if 
\begin{equation}
\label{eq: mass definition}
M^2=1
\quad
\rm{and}
\quad 
\{\alpha^{\ }_{1,2},M\}=0,
\end{equation}
yields
\begin{equation}
H^2
=
p^2_x\;+p^2_y\;+m^2.	
\end{equation}
Thus a mass matrix is one that
satisfies the relations in Eq.~(\ref{eq: mass definition})
and leads to the dispersion for a
massive Dirac particle $\varepsilon^{\ }_{\mathbf{p}} =\pm\sqrt{|\mathbf{p}|^2+m^2}$.

Let us then count the number of such matrices for cases of increasing
complexity, starting with the case of spinless fermions, marching
along to the 
most general, when we take into account
both spin and particle-hole gradings needed to discuss superconducting
graphene.

\subsection{Spinless case -- $4$-dimensional Dirac matrices}

Let us write the $4\times 4$ Dirac matrices that describe spinless
electrons in graphene in the Weyl representation:
\begin{equation}
\alpha^{\ }_i\equiv
\left(\begin{array}{cc} 
\tau^{\ }_i
&
0
\\
0
&
-
\tau^{\ }_i
\end{array}\right)
\equiv \sigma^{\ }_3\otimes\tau^{\ }_i
\;,\qquad
\beta
\equiv
\left(\begin{array}{cc} 
0
&
\tau^{\ }_0
\\
\tau^{\ }_0
&
0
\end{array}\right)
\equiv \sigma^{\ }_1\otimes\tau^{\ }_0
\;,
\end{equation}
where the $2\times2$ unit matrix $\tau^{\ }_{0}$ and the three Pauli
matrices $\tau^{\ }_{1}$, $\tau^{\ }_{2}$, and $\tau^{\ }_{3}$ act on
the sublattices indices ($\mathrm{A},\mathrm{B}$),
while the $2\times2$
unit matrix $\sigma^{\ }_{0}$ and the three Pauli matrices $\sigma^{\
}_{1}$, $\sigma^{\ }_{2}$, and $\sigma^{\ }_{3}$ act on the valley
indices ($+,-$). In this representation,
the four-component spinor is given by
$\Psi^\dagger\equiv
\left(\psi^\dagger_{+B},\psi^\dagger_{+A},\psi^\dagger_{-A},\psi^\dagger_{-B}\right)$.

We take the matrices $\alpha^{\ }_{1,2}$ to construct the kinetic part of
the Hamiltonian, such that the gapless case can be written as
\begin{equation}
{\cal K}^{\ }_0=p^{\ }_x\;\alpha^{\ }_1+p^{\ }_y\;\alpha^{\ }_2. 
\label{eq:def-K0}
\end{equation}
Now, the possible masses in the
4-dimensional representation of the Dirac Hamiltonian corresponds to
all matrices $M$ of the form
$X^{\ }_{\mu\nu}\equiv\sigma^{\ }_\mu\otimes\tau^{\ }_\nu$ (other than
$\alpha^{\ }_1\equiv X^{\ }_{31}$ and $\alpha^{\ }_2\equiv X^{\ }_{32}$) such that
$\{X^{\ }_{\mu\nu},\alpha^{\ }_{1}\}=0$ and $\{X^{\ }_{\mu\nu},\alpha^{\ }_{2}\}=0$. The
list of such matrices has 4 elements: 
$X^{\ }_{33}$,
$X^{\ }_{10}$,
$X^{\ }_{20}$,
and $X^{\ }_{03}$. These are then the 4 possible mass terms in spinless
graphene.

Physically, we can identify these four mass matrices as follows. One
perturbation that opens a gap $2\mu^{\ }_{\mathrm{s}}$ is a staggered
chemical potential, taking values $+\mu^{\ }_{\mathrm{s}}$ and
$-\mu^{\ }_{\mathrm{s}}$ in the two sublattices 
$\mathrm{A}$ and $\mathrm{B}$
of graphene. This is the mass term added by
Semenoff~\cite{Semenoff84}, 
and it corresponds to $X^{\ }_{33}$. A second
mass gap $2|\eta|$ arises by adding directed next-nearest-neighbor
hopping amplitudes in the presence of fluxes, but such that no net
magnetic flux threads a hexagonal Wigner-Seitz unit cell of graphene.
This perturbation, introduced by Haldane~\cite{Haldane88}, breaks
time-reversal symmetry (TRS) and corresponds to $X^{\ }_{03}$. Finally, a
real-valued modulation of the nearest-neighbor hopping amplitude with
a wave vector connecting the two Dirac points (i.e., a Kekul\'e
dimerization pattern for graphene) also opens a gap
$2|\Delta|$~\cite{Hou07}. This real-valued modulation of the
nearest-neighbor hoppings is parametrized by the complex order
parameter $\Delta=\mathrm{Re}\,\Delta+\mathrm{i}\,\mathrm{Im}\,\Delta$ whose
phase controls the angles of the dimerization pattern. This complex
order parameter translates into two real masses $\mathrm{Re}\,\Delta$
and $\mathrm{Im}\,\Delta$, corresponding respectively to $X^{\ }_{10}$ and
$X^{\ }_{20}$, bringing the total number of real-valued masses to four.

We can identify the microscopic origin of
these 4 masses according to the symmetries
that they respect (or break) 
at the level of the Dirac equation. 
To this end, let us look into two
symmetries that the Dirac Hamiltonian $H$ 
may or may not possess:

\begin{itemize}

\item Time-reversal symmetry (TRS) is satisfied when
\begin{equation}
X^{\ }_{11}\;H^*(-p)\;X^{\ }_{11}=H(p)
\;.
\label{eq:TRS4x4}
\end{equation}

\item Sub-lattice symmetry (SLS), 
also referred to as chiral symmetry in the literature,
is satisfied when
\begin{equation}
X^{\ }_{33}\;H(p)\;X^{\ }_{33}=-H(p)
\label{eq:SLS4x4}
\;.
\end{equation}
This symmetry operation corresponds to flipping the sign of the
wavefunction in one of the sublattices but not on the other (here, we
flip the sign on sublattice 
$\mathrm{A}$).

\end{itemize}

The Kekul\'e dimerization pattern corresponds to a spatial modulation
of the hopping matrix elements between sublattices 
$\mathrm{A}$ and $\mathrm{B}$, 
and
therefore changing the sign of the wavefunction on one sublattice
would reverse the overall sign of the Hamiltonian. And indeed, 
one verifies
that masses of the form $X^{\ }_{10}$ and $X^{\ }_{20}$
satisfy Eq.~(\ref{eq:SLS4x4}). The hopping amplitudes in the Kekul\'e
dimerization pattern are all real, and therefore they respect TRS,
which can also be checked via Eq.~(\ref{eq:TRS4x4}).

Both the staggered chemical potential or Semenoff mass $X^{\ }_{33}$ and the
Haldane mass $X^{\ }_{03}$ break SLS, as they involve couplings between
sites in the same sublattices. However, the staggered chemical
potential respects TRS, while the Haldane mass breaks it, as can be
checked using Eqs.~(\ref{eq:TRS4x4}) and (\ref{eq:SLS4x4}).

We shall introduce here a terminology that will be useful later on
when we consider systems with larger size representations, once spin
degrees of freedom and superconductivity are considered. For instance,
the Kekul\'e dimerization pattern corresponds to a valence-bond solid
(VBS) order parameter 
by
analogy to the terminology used for quantum
dimer models. A VBS order picks up a microscopic orientation that
translates into a complex-valued order parameter in the continuum
limit. Hence, we shall distinguish between the real (ReVBS) and
imaginary (ImVBS) parts of the VBS, and here they correspond to the
$\mathrm{Re}\,\Delta$ and $\mathrm{Im}\,\Delta$ of the Kekul\'e
distortion. Now, any mass matrix that does not satisfy the SLS
defined in
Eq.~(\ref{eq:SLS4x4}) corresponds to a microscopic order parameter for
which the fermion bilinear has the two lattice fermions sitting on the
same sublattice. This is the case of the mass associated to the
staggered chemical potential, which we may identify with a
charge-density wave (CDW). Finally, the Haldane mass implies a quantum
Hall effect (QHE). This nomenclature, along with the classification of
the 4 masses according to TRS and SLS, are summarized in
Table~\ref{tab:4-masses}.

\begin{table}
\center
\begin{tabular}{llll}
Mass matrix  & Order parameter & TRS & SLS \\
&&&\\ 
$X^{\ }_{10}$& {ReVBS} & {True} & {True} \\
$X^{\ }_{20}$& {ImVBS} & {True} & {True} \\
$X^{\ }_{33}$& {CDW}   & {True} & {False}\\
&&&\\ 
$X^{\ }_{03}$& {QHE}   & {False}& {False}
\end{tabular}
\vspace{.5cm}
\caption{
\label{tab:4-masses} 
The 4 mass matrices that can be added to the massless Dirac
Hamiltonian $\mathcal{K}^{\ }_{0}$ from Eq.~(\ref{eq:def-K0}) are of
the form
$X^{\ }_{\mu\nu}\equiv\sigma^{\ }_\mu\otimes\tau^{\ }_\nu$ and anticommute with
$\mathcal{K}^{\ }_{0}$. Each mass matrix can be assigned an order
parameter for the underlying microscopic model. Each mass matrix
preserves or breaks time-reversal symmetry (TRS), see
Eq.~(\ref{eq:TRS4x4}) and sublattice symmetry (SLS), see
Eq.~(\ref{eq:SLS4x4}).}
\end{table}

\subsection{Spinful case -- $8$-dimensional Dirac matrices}

Having warmed up with the $4\times4$ representations of the Dirac
matrices for the simpler spinless case, we now construct the
representations of all the masses for the case of graphene with the
spin degrees of freedom included, but still no superconductivity
considered.

To represent the single particle Hamiltonian $\mathcal{K}$, 
utilize the 64 8-dimensional Hermitean matrices
\begin{equation}
X^{\ }_{\mu^{\ }_{1}\mu^{\ }_{2}\mu^{\ }_{3}} \equiv 
{s}^{\ }_{\mu^{\ }_{1}}
\otimes
\sigma^{\ }_{\mu^{\ }_{2}}
\otimes
\tau^{\ }_{\mu^{\ }_{3}}
\label{eq: 64 matrices}
\end{equation}
where $\mu^{\ }_{1,2,3}=0,1,2,3$.  Here, we have introduced the three
families $ {s}^{\ }_{\mu^{\ }_{1}}$, ${\sigma}^{\ }_{\mu^{\ }_{2}}$,
and $ {\tau}^{\ }_{\mu^{\ }_{3}}$ of unit $2\times2$ and Pauli
matrices that encode the spin-1/2, valley, and sublattice degrees of
freedom of graphene, respectively.

The masses should be added to the massless Dirac Hamiltonian
\begin{equation}
{\cal K}^{\ }_0=p^{\ }_x\;\alpha^{\ }_1+p^{\ }_y\;\alpha^{\ }_2,
\qquad
{\rm where}\quad
\alpha^{\ }_1\equiv X^{\ }_{031}\quad
{\rm and}\quad
\alpha^{\ }_2\equiv X^{\ }_{032}.
\label{eq:def-K0-8}
\end{equation}
The possible masses in the 8-dimensional representation of the Dirac
Hamiltonian correspond to all matrices $M$ of the form
$X^{\ }_{\mu^{\ }_1\mu^{\ }_2\mu^{\ }_3}$ (other than $\alpha^{\ }_1$ and $\alpha^{\ }_2$) such
that $\{X^{\ }_{\mu^{\ }_1\mu^{\ }_2\mu^{\ }_3},\alpha^{\ }_{1}\}=0$ and
$\{X^{\ }_{\mu^{\ }_1\mu^{\ }_2\mu^{\ }_3},\alpha^{\ }_{2}\}=0$. One can carry out the
exercise of finding such matrices, obtaining the 16 matrices listed
in Table~\ref{tab:16-masses}.

We can classify these matrices according to the symmetries that they
respect (or break). The symmetries that the Hamiltonian $H$ may
possess are as follows:

\begin{itemize}

\item Time-reversal symmetry (TRS) is satisfied when
\begin{equation}
X^{\ }_{211}\;H^*(-p)\;X^{\ }_{211}=H(p)
\;.
\label{eq:TRS8x8}
\end{equation}

\item Sub-lattice symmetry (SLS) is satisfied when
\begin{equation}
X^{\ }_{033}\;H(p)\;X^{\ }_{033}=-H(p)
\label{eq:SLS8x8}
\;.
\end{equation}

\item Spin-rotation symmetry (SRS) is satisfied when
\begin{equation}
[X^{\ }_{100},H(p)]=[X^{\ }_{200},H(p)]=[X^{\ }_{300},H(p)]=0
\;.
\label{eq:SRS8x8}
\end{equation}

\end{itemize}

Notice that once we introduce spin degrees of freedom, we can 
decide
whether the Hamiltonian is spin rotational invariant or not. Also, we
can introduce terminology similar to the VBS and CDW cases we used in
labeling the masses for the spinless case. Masses that do not satisfy
the SLS defined in
Eq.~(\ref{eq:SLS8x8}) correspond to a microscopic order
parameters for which the fermion bilinear has the two lattice fermions
sitting on the same sublattice. Microscopic examples are, in addition
to the CDW already previously encountered, the spin-density waves
(SDW) such as N\'eel ordering, orbital currents leading to the quantum
Hall effect (QHE), and spin-orbit couplings leading to the quantum
spin Hall effect (QSHE). Whenever the instability can have a direction
associated to it in internal spin space, we add the
corresponding directional subscript $x,y,z$.

\begin{table*}
\center
\begin{tabular}{lllll}
Mass matrix  & Order parameter & TRS & SRS & SLS \\
&&&&\\ 
$X^{\ }_{010}$& {ReVBS} & {True} & {True} & {True} \\
$X^{\ }_{020}$& {ImVBS} & {True} & {True} & {True} \\
$X^{\ }_{033}$& {CDW}   & {True} & {True} & {False} \\
&&&&\\ 
$X^{\ }_{003}$& {QHE}   & {False}& {True} & {False}\\
&&&&\\ 
$X^{\ }_{110}$& {ReVBS}$^{\ }_{x}$& {False}& {False}& {True} \\
$X^{\ }_{210}$& {ReVBS}$^{\ }_{y}$& {False}& {False}& {True} \\
$X^{\ }_{310}$& {ReVBS}$^{\ }_{z}$& {False}& {False}& {True} \\
&&&&\\ 
$X^{\ }_{120}$& {ImVBS}$^{\ }_{x}$& {False}& {False}& {True} \\
$X^{\ }_{220}$& {ImVBS}$^{\ }_{y}$& {False}& {False}& {True} \\
$X^{\ }_{320}$& {ImVBS}$^{\ }_{z}$& {False}& {False}& {True} \\
&&&&\\ 
$X^{\ }_{103}$& {QSHE}$^{\ }_{x}$& {True} & {False}& {False} \\
$X^{\ }_{203}$& {QSHE}$^{\ }_{y}$& {True} & {False}& {False} \\
$X^{\ }_{303}$& {QSHE}$^{\ }_{z}$& {True} & {False}& {False} \\
&&&&\\ 
$X^{\ }_{133}$& {N\'eel}$^{\ }_{x}$& {False}& {False}& {False} \\
$X^{\ }_{233}$& {N\'eel}$^{\ }_{y}$& {False}& {False}& {False} \\
$X^{\ }_{333}$& {N\'eel}$^{\ }_{z}$& {False}& {False}& {False} \\
&&&&\\ 
\end{tabular}
\vspace{.5cm}
\caption{
\label{tab:16-masses} 
The 16 mass matrices that can be added to the massless Dirac
Hamiltonian $\mathcal{K}^{\ }_{0}$ from Eq.~(\ref{eq:def-K0-8}) are of
the form $X^{\ }_{\mu\nu\lambda}\equiv s^{\ }_{\mu} \otimes \sigma^{\ }_\nu \otimes \tau^{\ }_\lambda$ and anticommute with
$\mathcal{K}^{\ }_{0}$.
Each mass matrix can be assigned an order parameter for the
underlying microscopic model.
The latin subindex of the order parameter's name corresponds to the
preferred quantization axis in SU(2) spin space.
Each mass matrix preserves or breaks
time-reversal symmetry (TRS), see Eq.~(\ref{eq:TRS8x8}),
spin-rotation symmetry (SRS), see Eq.~(\ref{eq:SLS8x8}),
and sublattice symmetry (SLS), see Eq.~(\ref{eq:SRS8x8}).
        }
\end{table*}

The first 4 masses listed in Table~\ref{tab:16-masses} are physically
the same as those 4 listed in Table~\ref{tab:4-masses} for the
case of spinless electrons. These 4 masses correspond to order
parameters in the charge sector. The next 12 masses correspond to some
form of magnetic order. The simpler are the {N\'eel}$^{\ }_{x,y,z}$
order parameters along the three directions. The {N\'eel} states are
the SDW order associated to fermion bilinears at the same lattice
site. The {ReVBS}$^{\ }_{x,y,z}$ and {ImVBS}$^{\ }_{x,y,z}$ are 
their
counterparts
where the fermion bilinears are defined on the bonds instead of the
sites. Finally, the {QSHE}$^{\ }_{x,y,z}$ correspond to the quantum spin Hall
effect discussed by Kane and Mele in Ref.~\cite{Kane+Mele}.

\subsection{
Fully general case of single-layer graphene -- 
$16$-dimensional Dirac matrices
           }

To describe all symmetry-breaking instabilities with a local order
parameter in graphene we consider the Bogoliubov-de Gennes (BdG)
Hamiltonian
\begin{equation}
\hat{H}^{\ }_{\mathrm{BdG}}
=
\frac{1}{2}
\int d^{2}\boldsymbol{r}\,
\hat{\Psi}^{\dag}
\;H\;
\hat{\Psi}
\end{equation}
where $\hat{\Psi}$ is the 16-component Nambu spinor
\begin{equation}
\hat{\Psi} \equiv
\left(
\begin{array}{cccc}
\hat{\psi}^{\dag}_{\uparrow}, & 
\hat{\psi}^{\dag}_{\downarrow}, & 
\hat{\psi}^{\dag}_{\uparrow}, & 
\hat{\psi}^{\dag}_{\downarrow}
\end{array}
\right)^{\mathrm{t}}
\end{equation}
and $\hat{\psi}^{\ }_{s=\uparrow,\downarrow}$
is a 4-component fermion annihilation operator that accounts for
the 2 valley and the 2 sublattice degrees of freedom. 
The kernel of the BdG Hamiltonian has the block structure
\begin{equation}
H=
\left(\begin{array}{cc}
\mathcal{H}^{\ }_{\mathrm{p}\mathrm{p}}
& 
\mathcal{H}^{\ }_{\mathrm{p}\mathrm{h}}
\\
\mathcal{H}^{\dag}_{\mathrm{p}\mathrm{h}} 
& 
-\mathcal{H}^{\mathrm{t}}_{\mathrm{p}\mathrm{p}} 
\end{array}\right)
\label{eq: BdG block structure}
\end{equation}
where the
$8\times 8$ blocks
$\mathcal{H}^{\ }_{\mathrm{p}\mathrm{p}}$ 
and 
$\mathcal{H}^{\ }_{\mathrm{p}\mathrm{h}}$
act on the combined space of 
valley, sublattice, and spin degrees of freedom, 
and represent the normal and anomalous part of the BdG Hamiltonian,
respectively. These blocks satisfy
\begin{equation}
\mathcal{H}^{\dag}_{\mathrm{p}\mathrm{p}}=
\mathcal{H}^{\   }_{\mathrm{p}\mathrm{p}}
\quad
\mbox{(Hermiticity)},
\quad
\mathcal{H}^{\mathrm{t}}_{\mathrm{p}\mathrm{h}}=
-\mathcal{H}^{\ }_{\mathrm{p}\mathrm{h}}
\quad
\mbox{(Fermi statistics)}.
\label{eq: phs}
\end{equation}

To represent the single particle Hamiltonian $H$, 
define the 256 16-dimensional Hermitean matrices
\begin{equation}
X^{\ }_{\mu^{\ }_{1}\mu^{\ }_{2}\mu^{\ }_{3}\mu^{\ }_{4}} \equiv  
{\rho}^{\ }_{\mu^{\ }_{1}}
\otimes
{s}^{\ }_{\mu^{\ }_{2}}
\otimes
\sigma^{\ }_{\mu^{\ }_{3}}
\otimes
\tau^{\ }_{\mu^{\ }_{4}}
\label{eq: 256 matrices}
\end{equation}
where
$\mu^{\ }_{1,2,3,4}=0,1,2,3$.
Here, we have introduced the four families 
$  {\rho}^{\ }_{\mu^{\ }_{1}}$,
$     {s}^{\ }_{\mu^{\ }_{2}}$,
${\sigma}^{\ }_{\mu^{\ }_{3}}$,
and
$  {\tau}^{\ }_{\mu^{\ }_{4}}$
of unit $2\times2$ and Pauli matrices that encode
the particle-hole (Nambu), spin-1/2, valley, and sublattice 
degrees of freedom of graphene, respectively.

The Dirac kinetic energy $\mathcal{K}^{\ }_{0}$
of graphene
that accounts for the BdG block structure%
~(\ref{eq: BdG block structure})
is given by
\begin{equation}
{\cal K}^{\ }_0=p^{\ }_x\;\alpha^{\ }_1+p^{\ }_y\;\alpha^{\ }_2,
\quad
{\rm where}\quad
\alpha^{\ }_{1}\equiv X^{\ }_{0031} \quad
{\rm and}\quad
\alpha^{\ }_{2}\equiv X^{\ }_{3032}
\;.
\label{eq:def-K0-16}
\end{equation}

There are $64=4\times16$ mass matrices 
(i.e., $X^{\ }_{\mu^{\ }_{1}\mu^{\ }_{2}\mu^{\ }_{3}\mu^{\ }_{4}}$
that anticommutes with $\mathcal{K}^{\ }_{0}$).
Of these 64 mass matrices, only 36 satisfy the condition
\begin{equation}
X^{\ }_{1000}\;\;
X^{\mathrm{t}}_{\mu^{\ }_{1}\mu^{\ }_{2}\mu^{\ }_{3}\mu^{\ }_{4}}
\;\;
X^{\ }_{1000}
=
-
X^{\ }_{\mu^{\ }_{1}\mu^{\ }_{2}\mu^{\ }_{3}\mu^{\ }_{4}}
\label{eq: def PHS}
\end{equation}
for particle-hole symmetry (PHS)
and are thus compatible with the symmetry condition
$(\rho^{\ }_{1}\otimes s^{\ }_{0}\otimes \sigma^{\ }_{0}\otimes \tau^{\ }_{0}
  \hat{\Psi})^{\mathrm{t}}=\hat{\Psi}^{\dag}$
on the Nambu spinors 
[i.e., compatible with Eq.\ (\ref{eq: phs})].
All mass matrices with PHS are 
enumerated in Table \ref{tab: 36 masses}.

All 36 mass matrices from Table \ref{tab: 36 masses}
can be classified in terms of
the following (microscopic) 3 symmetry properties.

\begin{itemize}

\item A BdG Hamiltonian has time-reversal symmetry (TRS) when
\begin{equation}
X^{\ }_{0211}\,
H^*(-p)\,
X^{\ }_{0211}=
H(p) .
\label{eq: def TRS}
\end{equation}

\item A BdG Hamiltonian has
sublattice symmetry (SLS) when
\begin{equation}
X^{\ }_{0033}\,
H(p) \,
X^{\ }_{0033}=
-
H(p).
\label{eq: def SLS}
\end{equation}

\item A BdG Hamiltonian has SU(2) spin rotation symmetry (SRS) when
\begin{equation}
\left[ X^{\ }_{3100}, 
H(p)
\right]
=
\left[ X^{\ }_{0200}, 
H(p)
\right]
=
\left[ X^{\ }_{3300}, 
H(p)
\right]
=0.
\label{eq: def SRS}
\end{equation}

\end{itemize}

Identifying the physical meaning of the masses is done in a similar
way as explained in the simpler cases discussed previously (the
spinless and spinful cases without superconductivity). Below we
present the rational for the terminology in full generality.

The microscopic order parameter corresponding to a mass matrix
satisfying the SLS defined in Eq.~(\ref{eq: def SLS})
is a non-vanishing expectation
value for a fermion bilinear with the two lattice fermions residing on
the ends of a bond connecting sites in opposite sublattices.  We shall
say that such a mass matrix is associated to a valence-bond solid
(VBS) order parameter in analogy to the terminology used for quantum
dimer models. A VBS order picks up a microscopic orientation that
translates into a complex-valued order parameter in the continuum
limit. Hence, we shall distinguish between the real (ReVBS) and
imaginary (ImVBS) parts of the VBS. Triplet superconductivity 
(TSC) is also
possible on bonds connecting the two sublattices. The terminology TSC
will then also be used. To distinguish TSC with or without TRS we
shall reserve the prefixes Re and Im for real and imaginary parts.
This is a different convention for the use of the prefixes Re and Im
than for a VBS.

Any mass matrix that does not satisfy 
the SLS defined in Eq.~(\ref{eq: def SLS})
corresponds to a microscopic order parameter for which the fermion
bilinear has the two lattice fermions sitting on the same
sublattice. Microscopic examples are charge-density waves (CDW),
spin-density waves (SDW) such as N\'eel ordering, orbital currents
leading to the quantum Hall effect (QHE), spin-orbit couplings leading
to the quantum spin Hall effect (QSHE), singlet superconductivity
(SSC), or triplet superconductivity (TSC).

When SU(2) spin symmetry is broken by the order parameter, we add
a subindex $x$, $y$, or $z$ that specifies the relevant
quantization axis to the name of the mass matrix.
Moreover, TSC with SLS must be distinguished by the 2 possible
bond orientations (the underlying two-dimensional lattice has 2 
independent vectors connecting nearest-neighbor sites).
These 2 orientations are specified by the Pauli matrices used
in the valley and sublattice subspaces, 
i.e., by the 2 pairs of numbers 02 and 32.
Symmetry properties of all 36 PHS masses are 
summarized in Table \ref{tab: 36 masses}.

\begin{table*}
\center
\begin{tabular}{lllll}
Mass matrix  & Order parameter & TRS & SRS & SLS\\
&&&\\ 
$X^{\ }_{3010}$& {ReVBS} & {True} & {True} & {True}\\
$X^{\ }_{0020}$& {ImVBS} & {True} & {True} & {True} \\
$X^{\ }_{3033}$& {CDW}   & {True} & {True} & {False}\\
&&&\\ 
$X^{\ }_{3003}$& {QHE}   & {False}& {True} & {False}\\
&&&\\ 
$X^{\ }_{3110}$& {ReVBS}$^{\ }_{x}$& {False}& {False}& {True} \\
$X^{\ }_{0210}$& {ReVBS}$^{\ }_{y}$& {False}& {False}& {True} \\
$X^{\ }_{3310}$& {ReVBS}$^{\ }_{z}$& {False}& {False}& {True} \\
&&&\\ 
$X^{\ }_{0120}$& {ImVBS}$^{\ }_{x}$& {False}& {False}& {True} \\
$X^{\ }_{3220}$& {ImVBS}$^{\ }_{y}$& {False}& {False}& {True} \\
$X^{\ }_{0320}$& {ImVBS}$^{\ }_{z}$& {False}& {False}& {True} \\
&&&\\ 
$X^{\ }_{3103}$& {QSHE}$^{\ }_{x}$& {True} & {False}& {False}\\
$X^{\ }_{0203}$& {QSHE}$^{\ }_{y}$& {True} & {False}& {False}\\
$X^{\ }_{3303}$& {QSHE}$^{\ }_{z}$& {True} & {False}& {False}\\
&&&\\ 
$X^{\ }_{3133}$& {N\'eel}$^{\ }_{x}$& {False}& {False}& {False}\\
$X^{\ }_{0233}$& {N\'eel}$^{\ }_{y}$& {False}& {False}& {False}\\
$X^{\ }_{3333}$& {N\'eel}$^{\ }_{z}$& {False}& {False}& {False}\\
&&&\\ 
$X^{\ }_{2211}$& {ReSSC} & {True} & {True} & {False}\\
$X^{\ }_{1211}$& {ImSSC} & {False}& {True} & {False}\\
&&&\\ 
$X^{\ }_{1002}$& {ReTSC}$^{\ }_{02y}$& {True} & {False}& {True}\\
$X^{\ }_{2002}$& {ImTSC}$^{\ }_{02y}$& {False}& {False}& {True}\\
$X^{\ }_{1102}$& {ReTSC}$^{\ }_{02z}$& {False}& {False}& {True}\\
$X^{\ }_{2102}$& {ImTSC}$^{\ }_{02z}$& {True} & {False}& {True}\\
$X^{\ }_{1302}$& {ReTSC}$^{\ }_{02x}$& {False}& {False}& {True}\\
$X^{\ }_{2302}$& {ImTSC}$^{\ }_{02x}$& {True} & {False}& {True}\\
&&&\\ 
$X^{\ }_{1032}$& {ReTSC}$^{\ }_{32y}$& {False}& {False}& {True}\\
$X^{\ }_{2032}$& {ImTSC}$^{\ }_{32y}$& {True} & {False}& {True}\\
$X^{\ }_{1132}$& {ReTSC}$^{\ }_{32z}$& {True} & {False}& {True}\\
$X^{\ }_{2132}$& {ImTSC}$^{\ }_{32z}$& {False}& {False}& {True}\\
$X^{\ }_{1332}$& {ReTSC}$^{\ }_{32x}$& {True} & {False}& {True}\\
$X^{\ }_{2332}$& {ImTSC}$^{\ }_{32x}$& {False}& {False}& {True}\\
&&&\\ 
$X^{\ }_{1021}$& {ReTSC}$^{\ }_{y}$& {True} & {False}& {False}\\
$X^{\ }_{2021}$& {ImTSC}$^{\ }_{y}$& {False}& {False}& {False}\\
$X^{\ }_{1121}$& {ReTSC}$^{\ }_{z}$& {False}& {False}& {False}\\
$X^{\ }_{2121}$& {ImTSC}$^{\ }_{z}$& {True} & {False}& {False}\\
$X^{\ }_{1321}$& {ReTSC}$^{\ }_{x}$& {False}& {False}& {False}\\
$X^{\ }_{2321}$& {ImTSC}$^{\ }_{x}$& {True} & {False}& {False}\\
\end{tabular}
\vspace{.5cm}
\caption{
\label{tab: 36 masses} 
The 36 mass matrices with particle-hole symmetry (PHS),
see Eq.~(\ref{eq: def PHS}), 
for the massless Dirac Hamiltonian $\mathcal{K}^{\ }_{0}$
from Eq.~(\ref{eq:def-K0-16})
are of the form
~(\ref{eq: 256 matrices})
and anticommute with $\mathcal{K}^{\ }_{0}$.
Each mass matrix can be assigned an order parameter for graphene.
The latin subindex of the order parameter's name corresponds to the
preferred quantization axis in SU(2) spin space. The pair of
numeral subindices 02 and 32 are used to distinguish 
the two unit vectors spanning two-dimensional space.
Each mass matrix preserves or breaks
time-reversal symmetry (TRS), see Eq.~(\ref{eq: def TRS}),
spin-rotation symmetry (SRS), see Eq.~(\ref{eq: def SRS}),
and sublattice symmetry (SLS), see Eq.~(\ref{eq: def SLS}).
        }
\end{table*}

\section{Majorana bound states in superconducting graphene}
\label{sec:majoranas}

The unconventional relativistic-like band structure of graphene leads
to striking physical phenomena, for example when graphene
is placed in proximity with a superconductor. The problem of 
two dimensional Dirac fermions coupled to an s-wave superconductor 
was considered by Jackiw and Rossi~\cite{Jackiw81}, who have shown that the
fermionic spectrum displays a single zero energy mode
if the superconducting order parameter winds once about a given point in space
(the center of the vortex). This result is to be contrasted with the case of
non-relativistic s-wave superconductors,
for which no zero mode exists in the vortex core~\cite{Caroli64}. 

Because superconductivity mixes particles and holes, the second
quantized operator $\Gamma$ associated with the zero energy mode turns
out to be self-adjoint, that is to say, $\Gamma = \Gamma^{\dagger}$.
It is in the sense of being a ``real'' fermion that
a zero mode
represents
a condensed matter realization of Majorana
fermions~\cite{Majorana37,Wilczek2009}. An enormous amount of
interest rests upon Majorana states regarding their possible relevance
to constructing topological qubits: with two spatially
separated vortices, each of which hosting one zero mode, it is
possible to form a two level system (qubit) that stores the
information non-locally. The
parity of the number of zero modes per vortex core turns out to be
fundamental in determining the stability of the qubit. If, for
example, two zero modes exist at each vortex, generic perturbations
can split those modes apart causing the breaking down of the
stability. The general statement is then that an odd (even) number of
zero modes per vortex implies that one can form qubits that are stable
(unstable) against decoherence.

Let us suppose now that graphene is brought near to a
good conventional s-wave superconductor. By proximity
effect, electron-hole pairs can tunnel between graphene and the
superconductor. In this way, superconductivity can
be induced in graphene. In the presence of a superconducting vortex,
the low energy theory is an extension of that considered by Jackiw and Rossi. 
Zero mode solutions
exist~\cite{Ghaemi-Wilczek} but now there are
four Majorana modes per vortex, as opposed to just an isolated 
Majorana mode in the Jackiw-Rossi system. 
The even number of zero modes is a direct consequence
of the fermion doubling problem: in any 
TRS and local lattice theory, the number
of Dirac cones is even~\cite{Nielsen81}. Graphene can thus
be regarded, as far as the discussion of zero modes is concerned,
as four copies of the Jackiw-Rossi model.

For the purpose, at least as a question of principle,
of building topological qubits in graphene, or for
that matter in any lattice system with Dirac particles, it is
necessary to design a mechanism by which one can control the
parity of the number of zero modes bound to a singly-quantized vortex
in order to overcome the serious challenge originating from the
fermion doubling problem.

The same dilemma, but in a different context,
is present in the implementation of lattice gauge theories,
where the lattice regularization introduces spurious fermionic degrees 
of freedom. Wilson has proposed to overcome this problem
by introducing terms ``by hand'' in the
Hamiltonian with the effect of adding mass terms to those unwanted
fermions, thus removing them from the low energy sector~\cite{Wilson77}.
Although such Wilson masses indeed remove the extra
fermionic particles at tree-level, these perturbations have 
to be treated with great caution when quantum fluctuations are
taken into account. 

We advocate that for some lattice systems considered in 
condensed matter physics, the Wilson proposal is the 
way to control the parity of the number of
zero modes. Hereafter, we explain how the 
Wilson mechanism works
for graphene.

We recall that in graphene,
electrons with spin $\mathrm{s}=\uparrow,\downarrow$ hop on a honeycomb lattice that is made of two
triangular sublattices A and B. The conduction and valence
bands touch at the two non-equivalent points 
$\boldsymbol{K}^{\ }_{\pm}$ located at the opposite corners
in the hexagonal first Brillouin zone 
(see Ref.~\cite{graphene-review} for a review). 
Finally, to account for the
possibility of a SC instability, Nambu doublets are
introduced with the index p and h to distinguish particles
from their charge conjugate (holes). Hence, after linearization
of the spectrum about the Fermi points
$\boldsymbol{K}^{\ }_{\pm}$,
this leads to a single-particle
kinetic energy represented by a 16$\times$16 dimensional matrix
$
\mathcal{K}^{\ }_0\equiv
p^{\ }_{x}\,\alpha^{\ }_{1}
+
p^{\ }_{y}\,\alpha^{\ }_{2}
$.
Here, $\alpha^{\ }_{1}$ and $\alpha^{\ }_{2}$ are two
16$\times$16 dimensional Dirac matrices.
   
As discussed in Sec.~\ref{sec:masses} above (and in
Ref.~\cite{Ryu09}), there exists 36 distinct order parameters (listed
in Table \ref{tab: 36 masses}) such that any one, when added to
$\mathcal{K}^{\ }_0$, opens a spectral gap. These order parameters were
identified by seeking all 16$\times$16 matrices from the Clifford
algebra that anticommute with $\mathcal{K}^{\ }_0$.  Among these order
parameters, two (a real and an imaginary part) correspond to one
complex-valued order parameter that is associated with singlet
superconductivity. We shall denote the two corresponding 16$\times$16
matrices from the Clifford algebra by $M^{\ }_{\mathrm{ReSSC}}$ and
$M^{\ }_{\mathrm{ImSSC}}$ and define the perturbation $ \mathcal{H}^{\
}_{\Delta}\equiv \Delta^{\ }_{1} M^{\ }_{\mathrm{ReSSC}} + \Delta^{\
}_{2} M^{\ }_{\mathrm{ImSSC}} $ that opens the spectral gap
$2|\Delta|$ with the complex-valued $\Delta\equiv \Delta^{\
}_{1}+\mathrm{i}\,\Delta^{\ }_{2}$ parametrized by the real-valued $\Delta^{\
}_{1}$ and $\Delta^{\ }_{2}$ when added to $\mathcal{K}^{\ }_0$.

Next, we would like to select other order parameters among the
remaining 34(=$36-2$) masses that 
{\it compete} with superconductivity,
i.e., masses that do not add in quadrature with the two
superconducting masses $M^{\ }_{\mathrm{ReSSC}}$ and
$M^{\ }_{\mathrm{ImSSC}}$. Matrices corresponding to masses that add in
quadrature anticommute, while matrices corresponding to masses that
compete commute. We thus seek all 16$\times$16 matrices from the
Clifford algebra that (i) anticommute with $\mathcal{K}^{\ }_0$ and (ii)
commute with $\mathcal{H}^{\ }_{\Delta}$. In this way, we find all 10
TRS-breaking order parameters listed in Table~\ref{table-1} that alone
would open a gap in the Dirac spectrum if not for their competition
with the gap induced by singlet superconductivity.  Within this set of
10 matrices one can form groups of at most $4$ matrices that are
mutually commuting and therefore can be simultaneously
diagonalized. Let us choose the $4$-tuplet $\{\mathrm{ReVBS}^{\ }_{x},
\mathrm{ImVBS}^{\ }_{y}, \mathrm{Neel}^{\ }_{z}, \mathrm{IQHE} \}$ for
concreteness, but the results hereafter apply to any other such
$4$-tuplet of commuting mass matrices among the set of 10.  
Observe that any member of this
 $4$-tuplet breaks TRS. This property will allow us to overcome
the fermion doubling barrier~\cite{Nielsen81}.
It is then
possible to show that

\begin{eqnarray}
\mathcal{H}&=& 
\boldsymbol{p} 
\cdot 
\boldsymbol{\alpha} 
+
\Delta^{\ }_{1}\, M^{\ }_{\mathrm{ReSSC}}
+ 
\Delta^{\ }_{2}\, M^{\ }_{\mathrm{ImSSC}}
\nonumber\\
&& {}
+ 
m^{\ }_{1}\, M^{\ }_{\mathrm{ReVBS}^{\ }_{x}}
+ 
m^{\ }_{2}\, M^{\ }_{\mathrm{ImVBS}^{\ }_{y}}
+ m^{\ }_3\, M^{\ }_{\mathrm{Neel}^{\ }_{z}} 
+ 
\eta\, M^{\ }_\mathrm{IQHE},
\end{eqnarray}
after a unitary transformation, can be
brought into the form

\begin{equation}
\mathcal{H}\equiv
\left(\begin{array}{cccc}
\mathcal{H}^{\ }_{1}
&0
&0
&0\\
0
&\mathcal{H}^{\ }_{2}
&0
&0\\
0
&0
&\mathcal{H}^{\ }_{3}
&0\\
0
&0
&0
&\mathcal{H}^{\ }_{4}
\end{array}\right)
\label{eq: continuum limit noncentrosymmetric a}
\end{equation}
with the 4$\times$4 Hermitean blocks
\begin{equation}
\mathcal{H}^{\ }_{j} = 
\left(\begin{array}{cccc}
-\eta^{\ }_{j} 
& 
p
& 
\delta^{\ }_{j} 
& 
0 
\\
\overline{p}
& 
\eta^{\ }_{j} 
& 
0 
& 
\delta^{\ }_{j} 
\\
\overline{\delta^{\ }_{j}} 
& 
0 
& 
-\eta^{\ }_{j} 
& 
-p
\\
0 
& 
\overline{\delta^{\ }_{j}} 
& 
-\overline{p}
& 
\eta^{\ }_{j}
\end{array}\right)
\label{eq: continuum limit noncentrosymmetric b}
\end{equation}
whereby the complex notation $p\equiv p^{\ }_{1}+\mathrm{i}\,p^{\ }_{2}$ 
is used for the momenta and
$\overline{x}$ denotes the complex conjugate of $x$.
It is found that
\begin{eqnarray}
&&
\delta^{\ }_{1,2,3,4}\equiv\Delta,
\\
&&
\eta_1\equiv 
-m^{\ }_1+m^{\ }_2+m^{\ }_3+\eta,
\\
&&
\eta_2\equiv\, 
m^{\ }_1-m^{\ }_2+m^{\ }_3+\eta,
\\
&&
\eta_3\equiv\, 
m^{\ }_1+m^{\ }_2-m^{\ }_3+\eta,
\\
&&
\eta_4\equiv 
-m^{\ }_1-m^{\ }_2-m^{\ }_3+\eta.
\end{eqnarray}

The breaking of the $16$-dimensional matrix into four independent
sectors is key to the ability of controlling the number of zero modes.
The argument goes as follows.  If all the $\eta^{\ }_{j}$'s are
zero and the SC order parameter has a single vortex with unit winding
number, there are 4 Majorana fermions bound to
it~\cite{Ghaemi-Wilczek}. Indeed, in this limit one has precisely four
copies of the Jackiw-Rossi model, with any
one copy delivering one zero-mode.

However, as the magnitudes of the $|\eta^{\ }_{j}|$'s increase, there will
be a phase transition every time that $|\eta^{\ }_{j}|=|\Delta(r=\infty)|$,
where $|\Delta(r=\infty)|$ is the bulk value of the order parameter
far away from the center of the vortex. There is no zero-mode attached
to vortices in the $j$-th copy when
$|\eta^{\ }_{j}|>|\Delta(r=\infty)|$,
as can be explicitly checked.
Indeed, this gapped phase is adiabatically connected to the limit
$|\Delta(r=\infty)|/|\eta^{\ }_{j}|=0$
with no superconductivity, i.e., no support for a zero mode
(in this gapped phase, the spectral symmetry of the BdG
Hamiltonian prevents any change of the parity in the number of zero modes).
Therefore, one can knock out the Majorana fermions one
by one by taking the values of the four $|\eta^{\ }_{j}|$'s across the
phase transitions.

In summary, we have identified a mechanism to overcome the
fermion-doubling barrier that can prevent the attachment of an odd
number of Majorana fermions to the core of SC vortices in
graphene-like tight-binding models. This mechanism relies on a
topological charge that measures the parity in the number of Majorana
fermions attached to an isolated vortex and the use of TRS-breaking
order parameters that compete with each other and with the SC order
parameter to knock out one by one the Majorana fermions. Therefore one
can selectively switch between even and odd numbers.

\begin{table}
\center
\begin{tabular}{lllll}
Mass matrix &
Order parameter & 
TRS & 
SRS & 
SLS\\
&&&&\\ 
$X^{\ }_{3003}$&
{IQHE} & 
{False}& 
{True} & 
{False}
\\
&&&&\\ 
$X^{\ }_{3110}$&
{ReVBS}$^{\ }_{x}$& 
{False}& 
{False}& 
{True} 
\\
$X^{\ }_{0210}$&
{ReVBS}$^{\ }_{y}$& 
{False}& 
{False}& 
{True} 
\\
$X^{\ }_{3310}$&
{ReVBS}$^{\ }_{z}$& 
{False}& 
{False}& 
{True} 
\\
&&&&\\ 
$X^{\ }_{0120}$&
{ImVBS}$^{\ }_{x}$& 
{False}& 
{False}& 
{True} 
\\
$X^{\ }_{3220}$&
{ImVBS}$^{\ }_{y}$& 
{False}& 
{False}& 
{True} 
\\
$X^{\ }_{0320}$&
{ImVBS}$^{\ }_{z}$& 
{False}& 
{False}& 
{True} 
\\
&&&&\\ 
$X^{\ }_{3133}$&
{N\'eel}$^{\ }_{x}$& 
{False}& 
{False}& 
{False}
\\
$X^{\ }_{0233}$&
{N\'eel}$^{\ }_{y}$& 
{False}& 
{False}& 
{False}
\\
$X^{\ }_{3333}$&
{N\'eel}$^{\ }_{z}$& 
{False}& 
{False}& 
{False}
\end{tabular}
\caption{ 
\label{table-1}
The 10 mass matrices with particle-hole symmetry (PHS) that
anticommute with $\alpha^{\ }_{1}$ and $\alpha^{\ }_{2}$ and
commute with the singlet SC masses 
$M^{\ }_{\mathrm{ReSSC}}$
and 
$M^{\ }_{\mathrm{ImSSC}}$.  Each mass matrix can be
assigned an order parameter for the underlying microscopic model. 
The latin subindex of the order parameter's name corresponds to the
preferred quantization axis in SU(2) spin space. Each mass matrix
either preserves or breaks time-reversal symmetry (TRS), spin-rotation
symmetry (SRS), and sublattice symmetry (SLS). An explicit 
representation defined in Ref.~\cite{Ryu09} is given in the last
column.}
\end{table}


\vspace{2cm}

\begin{thebibliography}{99}

\bibitem{Novoselov04}
  K.S. Novoselov, A. K. Geim, S. V. Morozov, D. Jiang, S. V. Dubonos,
  I. V. Grigorieva, and A. A. Firsov,
Science \textbf{306}, 666 (2004).

\bibitem{graphene-review}
A. H. Castro Neto, F. Guinea, N. M. Peres, K. S. Novoselov, and A. K. Geim, 
Rev.\ Mod.\ Phys.\ \textbf{81}, 109 (2009).

\bibitem{Wallace47}
P. R. Wallace,
Phys.\ Rev.\ \textbf{71}, 622 (1947).

\bibitem{Jackiw81}
R. Jackiw and P. Rossi, 
Nucl.\ Phys.\ B \textbf{190}, 681 (1981).

\bibitem{Hou07}
C.-Y. Hou, C. Chamon, and C. Mudry, 
Phys.\ Rev.\ Lett.\ \textbf{98}, 186809 (2007).

\bibitem{Ghaemi-Wilczek}
P. Ghaemi and F. Wilczek, 
arXiv:0709.2626 (unpublished).

\bibitem{Fu-Kane}
L.\ Fu, C.\ L.\ Kane,
Phys.\ Rev.\ Lett.\ \textbf{100}, 096407 (2008).


\bibitem{Ryu09}
S. Ryu,C. Mudry, C.-Y. Hou, and C. Chamon, 
Phys.\ Rev.\ B \textbf{80}, 205319 (2009). 

\bibitem{Santos10}
L. Santos, S. Ryu, C. Chamon, and C. Mudry, 
Phys.\ Rev.\ B \textbf{82}, 165101 (2010). 

\bibitem{Semenoff84} 
G. W. Semenoff, 
Phys.\ Rev.\ Lett.\ \textbf{53}, 2449 (1984).

\bibitem{Haldane88} 
F. D. M. Haldane, 
Phys.\ Rev.\ Lett.\ \textbf{61}, 2015 (1988).

\bibitem{Kane+Mele}
C. L. Kane and E. J. Mele , 
Phys.\ Rev.\ Lett.\ \textbf{95}, 226801 (2005).

\bibitem{Nielsen81}
H. B. Nielsen and M. Ninomiya,
Nucl.\ Phys.\ B \textbf{185}, 20 (1981);
\textbf{193}, 173 (1981).

\bibitem{Caroli64}
C. Caroli, P.G. De Gennes, and J. Matricon,
Phys.\ Lett.\ \textbf{9}, 307 (1964).

\bibitem{Majorana37}
Majorana, E.
Nuovo Cimento \textbf{5}, 171 (1937).

\bibitem{Wilczek2009}
F. Wilczek,
Nature Physics \textbf{5}, 614 (2009).

\bibitem{Wilson77} 
K. Wilson in 
\textit{New Phenomena in Subnuclear Physics}, 
Edited by A. Zichichi (Plenum, New York, 1977).

\end{thebibliography}
\end{document}